\documentclass[10pt,journal,compsoc]{IEEEtran}

\usepackage{url}
\usepackage{graphicx}
\usepackage{cite}
\usepackage{subfigure}
\usepackage{wrapfig}
\usepackage{tikz}
\usetikzlibrary{arrows,shapes, fit, 
decorations.pathreplacing,positioning, arrows.meta,shapes.arrows,chains}

\usepackage{amsmath}

\usepackage{algorithm}
\usepackage{algorithmicx}
\usepackage{algpseudocode}

\algrenewcommand{\alglinenumber}[1]{\algprintlinenumber{#1}:}
\algnewcommand{\algprintlinenumber}[1]{\footnotesize#1}
\newcommand{\myref}[1]{\hyperref[#1]{\algprintlinenumber{\ref*{#1}}}}
\algdef{SE}[DOWHILE]{Do}{doWhile}{\algorithmicdo}[1]{\algorithmicwhile\ #1}%
\newcommand{\algosizefinal}{\small}

\newcommand{\CAS}{{\tt Compare\&Swap}}
\newcommand{\slock}{{\tt spn\_obj}} 
\newcommand{\sws}{{\tt sws}} 
\newcommand{\thc}{{\tt thc}} 
\newcommand{\wuc}{{\tt wuc}} 
\newcommand{\futex}{{\tt slp\_obj}} 
\newcommand{\lstate}{{\tt lstate}}

\newcommand{\remove}[1]{}

\begin{document}
\author{
Romolo Marotta, Davide Tiriticco, Pierangelo Di Sanzo, Alessandro Pellegrini, Bruno Ciciani, \linebreak Francesco Quaglia
}

\title{Mutable Locks:\\Combining the Best of Spin and Sleep Locks}

\IEEEtitleabstractindextext{
\begin{abstract}
In this article we present Mutable Locks, a synchronization construct with the same execution semantic of traditional locks (such as spin locks or sleep locks), but with a self-tuned optimized trade off between responsiveness---in the access to a just released critical section---and CPU-time usage during threads' wait phases. 
It tackles the need for modern synchronization supports, 
in the era of multi-core machines, 
whose runtime behavior should be optimized along multiple dimensions (performance vs resource consumption) with no intervention by the application programmer. 
Our proposal is intended for exploitation in generic concurrent applications where scarce or none knowledge is available about the underlying software/hardware stack and the actual workload, an adverse scenario for static choices between spinning and sleeping faced by mutable locks just thanks to their hybrid waiting phases and self-tuning capabilities. 
%which has lead to the need for systematically synchroninizing threads running critical sections that can be of %arbitrary (hence also minimal) length.
%\keywords{thread synchronization, multi-core machines} 
%, perfomance vs resource usage}
\end{abstract}

}

\maketitle

\section{Introduction}
\label{introduction}

\IEEEPARstart{M}{odern} multi-core chipsets and the ever-growing adoption of concurrent programming in daily-usage software are posing %thread 
new synchronization challenges.
 %on how thread synchronization 
 %in the access to shared data 
 %should occur.
 %, because of both performance end energy efficiency. 
% On the hardware side, n
Non-coherent cache architectures---such as Intel SCC (Single-chip Cloud Computer) \cite{intelscc}---look to be a way for reducing the complexity of classical cache-coherency protocols. However, these solutions significantly impact software design, since hardware-level coherency becomes fully demanded from software---which needs to rely on explicit message passing across cores to let updated data values flow in the caching hierarchy. 
 
 Hardware-Transactional-Memory (HTM) 
 %is emerging in modern processors---like the Intel Haswell family---as a support for 
 allows atomic and isolated accesses to slices of shared data in multi-core machines. This solution is however not viable in many scenarios because of limitations in the HTM firmware,
 %has a set of limitations,
 like the impossibility to successfully finalize (commit) data manipulations in face of hardware events such as interrupts.
 %and mode chances. 
 For this reason, HTM is often used in combination with traditional {\em locking primitives}, which enable isolated shared-data accesses otherwise not sustainable via HTM.
 
 The Software Transactional Memory (STM) counterpart avoids HTM-related limitations.
%---while jointly trying to hide synchronization complexities from the specific application. 
However, STM internal mechanisms (e.g. \cite{Dice06}) still rely on locks to enable atomic and isolated management of the metadata that the STM layer exploits to assess the correctness (e.g. the isolation) of data accesses performed by threads.
%the transactions encapsulating the application logic. 
Furthermore, locking is still exploited as a core mechanism in software-based shared-data management approaches for multi-core machines like Read-Copy-Update (RCU) \cite{mckenney98}, where readers are allowed to concurrently access shared data with respect to writers, but concurrent writers are anyhow serialized via the explicit usage of locks.

Despite the rising trend towards 
%the construction of 
differentiated synchronization supports,
%for shared-data (or resource) access, 
locking still stands as a core synchronization mechanism.
%that is more or less explicitly exploited in such rising supports. 
%in scenarios were concurrent threads access shared data (or resources) on milt-core machines. 
Therefore, optimizing the runtime behavior of locking primitives is a core achievement for software operations carried out on nowadays multi-core hardware.

Actually, we can distinguish among two main categories of locks: 1) \emph{spin locks}, which are based on threads actively waiting for the ownership in the access to the targeted shared resource;
%---for executing the so-called critical section; 
2) \emph{sleep locks}, which make threads not run on any CPU-core until they can (retry to) acquire the access ownership.
%in the access to the target shared-data slice.
As well known, the first category can operate in user-space, while the latter requires the interaction with and the support of the underlying Operating System (OS) kernel.
%In fact, going to sleep, i.e. being suspended, requires that the OS does not schedule sleeping threads.

%\subsubsection{/*Problem statement*/}

Spin locks are often preferred in HPC applications---where low or none time-sharing interference in spinning phases (or more generally in the usage of CPU by threads) is expected---since they ensure the lowest latency while acquiring the ownership of the lock.
However, they do not care about metrics such as CPU usage. 
In fact, CPU cycles wasted in spinning operations because of conflicting accesses to critical sections may result non-negligible at non-minimal thread counts.
Moreover, this might increase the impact of hardware contention on performance, since spin operations typically involve atomic instructions that trigger the cache-coherence firmware.
%Consequently, resorting to sleeping could be one possible strategy (the one explored in this work) to face this new challenge.
Conversely, sleep locks save CPU cycles and reduce hardware contention, thus representing the obvious alternative to spin locks when resource usage is a concern. 
However, they might increase the latency in the access to a critical section because of delays introduced by the OS while awakening and scheduling threads.

In this scenario, developers might benefit of a lock supporting both active and passive waiting phases and able to determine the best choice between spinning and sleeping at run time.
Such an approach can guarantee that non-predicted changes in the workload cannot hamper the overall system performance and can relieve developers from taking static choices between spinning and sleeping, which might be always inadequate with a dynamic workload, reducing the experimental evaluation and development time.

%waking up threads and making them sleep (as well as reduce cache locality).
%Moreover, if threads are not awake when the lock has been released, we might experience an incresed latency due to the context-switching.

To tackle the limitations of spin and sleep locks, in this article we present a new synchronization support called {\em mutable lock}. Our solution is based on a non-trivial combination of spin and sleep primitives, which gives rise to a state machine driving the evolution of threads in such a way that sleep-to-spin transitions are envisaged as a means to always guarantee that some thread is already awaken when the critical section is released. 
Hence, it can access the critical section with no additional delay caused by the OS awakening phase. On the other hand, the sleep phase is retained as a means for controlling the waste of resources that would otherwise be experienced with pure spin locking.

%We also note that our approach is definitely different from the trivial combination of spinning and sleeping already offered by a few implementations of locking primitives, such as mutexes in the {\sf glibc} {\tt pthread} library. These combinations are simply based on an initial attempt to spin for a while, and then on resorting to sleep. 
%In this solution there is no will to 
%So they do not optimize the number of spinning threads and there is no sleep-to-spin transition---anticipating the OS %awake of threads with respect to the release of the critical section by another thread---which we instead exploit as a %means for promptness in the access to the critical section.

Our mutable locks ship with the support for the autonomic tuning of the transitions between sleep and spin phases---or the choice of one of the two upon the initial attempt to access the critical section---which is implemented as a control algorithm encapsulated into the locking primitives. Furthermore, our solution is fully transparent, and can be exploited by simply redirecting the API of the  locking primitive originally used by the programmer to our mutable locks library\footnote{Code available at \url{https://github.com/HPDCS/libmutlock}}.

%For these reasons, we have designed a lock that combines the strenghts of spin locks and sleep locks in terms of responsiveness and clock-cycles savings.
%This is achieved by introducing the \emph{spinning-window} mechanism, which allows to control the number of waiting threads enabled to spin.
%Such an approach is able to ensure that i) there is always the minimum number of active threads for accessing the critical section with almost no latency and ii) clock-cycles consumption is reduced since all other threads are sleeping. 
%Clearly, this is an indirect approach towards energy effiency.
%In fact, we do not tackle the problem by relying on instructions with a low energy profile or manipulating CPU frequencies, but by saving clock cycles that can be spent by other application in the system improving its overall throughput.

%\subsubsection{/*Contributions*/}

%Our work has contributed to the design of a mutable lock able to:
%\begin{itemize}
%\item combine spinning and sleeping waiting phases;
%\item control the number of threads enabled to spin;
%\item can be customized with user defined policies;
%\item can guarantee starvation freedom according to the chosen underlying synchronization objects.
%\end{itemize} 

%\subsubsection{Paper Organization}

The remainder of this article is structured as follows. 
%Section 2 presents some useful background information about locks. 
In Section \ref{related} we discuss related work. Section \ref{mutable} presents the design of our mutable locks. Experimental results for a comparison with other conventional lock implementations are reported in Section \ref{data}. Section 5 gives the conclusions.

\section{Related Work}
\label{related}

Spin locks have been originally implemented by only relying on atomic read/write instructions
\cite{Dijkstra:1965:SPC:365559.365617,Lamport:1974:NSD:361082.361093}.
%, such as .
However, this solution had limited applicability to scenarios where the number of threads to synchronize was known at compile/initialization time and could not change at runtime. Such a limitation was overcome by recurring to atomic Read-Modify-Write (RMW) instructions, like 
%as \TAS\ (denoted as TAS)
%, \FAD\ (FAD) 
%and 
 \CAS\ (CAS) in modern processors. The main idea behind RMW-based spin locks is the one of repeatedly trying to atomically switch a variable from a value to another value---the so-called test-and-set operation.
%variable to a specific value.
If a thread succeeds in this operation, it can proceed and execute the critical section, otherwise it has to continuosly retry the operation---this is the spin phase.
%, namely spinning. 
%Finally, when a thread completes the critical section, it releases the lock by setting the variable to a specific value. 
%As noted in Section \ref{introduction}, s
Spin locks are greedy in terms of clock cycles usage, thus leading to non-minimal waste of resources in scenarios with non-negligible likelihood of thread conflict in the access to critical sections. This problem is further exacerbated by the fact that RMW instructions make intensive usage of state transitions in the hardware-level cache coherency protocol. This in turn can impact the cache access latency by other threads, including the one that is currently owning the critical section. Clearly, the more threads spin at the same time, the worse the scenario becomes.

%large usage of 

%On the other hand, 

%Spin locks adopting the above mentioned scheme are typically called \emph{test-and-set} spin locks and,
%The TAS instruction atomically exchanges the value of a memory cell with a supplied value and returns the old content.
%A simple test-and-set lock can be implemented as a unique boolean initially set to false, where the challenge phase consists in repeatedly updating such boolean as true with a TAS.
%The winner thread is the unique one that reads false as previous value, while others continue to update the lock variable.
%The winner can proceed executing the critical section and, once completed, it releases the lock by setting the lock variable to false.
%At this point, one of the challenging threads will read the old value as false and the computation will go on.
%unfortunately,
%this simple spin lock implementation
%do not perform well in a real machine, since %typically 
%they are very greedy in executing atomic instructions 
%involving the cache coherence protocol very frequently. % that is quite expensive.
%Moreover, such significant amount of memory traffic makes threads running the critical section slower, impacting the overall system %performance.

%The literature proposes spin-lock impementitons overcoming a few  of the the limitations of simple test-and-set spin locks.
%In particular,   
%A significant improvement on the test-and-set spin lock is the 

The \emph{test-and}-test-and-set spin lock \cite{Rudolph1984} makes
%More precisely, 
challenging threads 
%, instead of continuously trying to acquire the lock with a TAS, they 
continuously check the lock variable until it is released %assumes a false value
and, only in this case, they try to acquire it via RMW instructions. % perform a TAS.
This allows threads to \emph{spin} (read the actual value of the lock variable) in cache without disturbing others, thus generating cache/memory traffic only when strictly needed.
%---signaling the release/acquisition of the lock.

The authors of \cite{Anderson:1990:PSL:628891.628973} introduce %propose % another strategy to resolve the memory-traffic issue of test-and-set locks. 
%It consists in 
%the introduction of
a simple back-off time before attempting to re-acquire the lock.
% executing a TAS.
Anyhow, such a strategy requires some variables to be set up, such as the maximum and minimum back off time, that cannot be universal across any hardware architecture and/or workload \cite{Scott2013}.

%Even if test-and-set with back off and test-and-test-and-set locks perform well in general cases, they might lead to starvation. 
Spin locks might lead to starvation since
%In fact, 
there is no assurance that a given thread wins the challenge eventually.
%Such a limitation is overcome with ticket locks \cite{Mellor-Crummey1991}, that are essentially two integer variables
%(one storing the served ticket, the other storing the next available ticket) updated with a \FAD\ (FAD) RMW %instruction, which atomically increments the value of a memory cell and returns its value before the update. However, %these variables are repeatedly accessed during spin operations leading again to contention on the caching system.
%
%One variable $A$ stores the actual served ticket and the other ($B$) the next available ticket. 
%A new arriving thread executes a \FAD\ on $B$ obtaining the old value (the ticket) and incrementing its value and spins (with optional back off) until $A$ become equal to $B$.
%When this condition holds, the thread has acquired the lock and can continue its computation.
%When it wants to release the lock, it has to set the value of $A$ to its ticket plus one.
%At this point, a unique spinning thread acquires the lock.
%
%The ticket lock overcomes the starvation problem of classical spin locks implementation, but it shares with them the same drawback: an almost %unique synchronization variable.
%In fact, such variable, regardless of if it is a bit or a counter, is repeatedly read by threads during spin operations, creating a %contention at the lowest level of cache at least.
%This might induce thrashing with the core holding the lock and reducing the overall system performance, since the execution time for the %critical section might be elongated.
The authors of \cite{Mellor-Crummey1991} introduce the queued spin lock to resolve this issue.
%A queued spin lock
It is a linked list where the first connected node is owned by the thread holding the lock, while others are inserted in a FIFO order by threads trying to access the critical section.
Such threads spin on a boolean variable encapsulated in their individual nodes.
This guarantees that each spinning thread repeatedly reads a memory cell different from other threads and a releasing thread updates a cacheline owned by a unique CPU-core, which significantly reduce the pressure on the cache management firmware. 
%This ensures an increased scalability with respect to the number of cores in a single machine even in a Non-Uniform Memory Access (NUMA) %configuration.

In all the above solutions, there is no direct attempt to control the number of threads spinning at each time instant, as instead we do in our mutable locks thanks to the smart combination of spin/sleep phases and sleep-to-spin transitions. Clearly, such a limitation of the literature approaches can lead to catastrophic consequences on performance and resource usage when there is a relevant hardware contention, e.g. when applications are executed with more threads than cores.
%a number of threads $T$, on a machine with $N$ cores, such that $N<T$. 
%In this scenario we can expect that each thread is CPU-scheduled after $N/T$ time slots. Hence, if the critical %section is (much) longer than the assigned time slot, the wall-clock time required for completing the critical section increases with the number of threads.
Furthermore, when recurring to FIFO locks 
%(e.g. ticket and queued locks) 
an anti pattern emerges.
The FIFO semantics imposes that, when a thread releases the lock, \emph{one specific thread} has to acquire it---the one standing at the head of the FIFO queue.
It follows that delays (for example a CPU descheduling) affecting a thread impact all its successors in the queue. 
%---but, as stated before, a given thread is rescheduled after $N/T$ time slots.
This increases considerably the residence time (queue time plus critical section execution time) and consequently the overall system performance can be impaired.
%Conversely, test-and-set locks do not suffer from this anti pattern since \emph{any thread} winner of the ``challenge'' can acquire the lock.
As overall considerations, spin locks are typically avoided with long critical sections just because of the above motivations, but they do anyhow suffer from the problem of waste of resources in face of conflicting accesses to the critical section. Our mutable locks cope with both these problems, since the smart combination between spin and sleep phases avoids the antipattern where a thread running a long critical section is descheduled in favor of one simply spinning for the access to the critical section.

As hinted, sleep locks---based on OS blocking services---represent the opposite solution to synchronization, and are aimed at avoiding usage of resources (that would take place with spin locks) during wait phases preceding the access to the critical section. OS implementations offer sleep locks since their very beginning, and various improvements in these synchronization constructs have been devised in order to enable flexible synchronization schemes, involving awake conditions resulting as the combination of the state of multiple sleep locks. Examples are the System V semaphores offered by Posix \cite{posix} or the wait-for-multiple-object primitive offered by  WinAPI \cite{winapi}. In any case, all the sleep locks based on blocking OS services share the common drawback that, as soon as a critical section is released, there is no guarantee that a thread willing to access the critical section is already CPU-dispatchable (or already dispatched). In fact, it might have gone sleeping, thus needing to undergo a wake-up phase bringing it back onto the OS run-queue. Overall, we may experience a delay in the access by this thread to the critical section, which in turn may hamper performance, especially when the critical section is short---a 
%. Again, this 
problem 
%is 
exacerbated at higher concurrency.

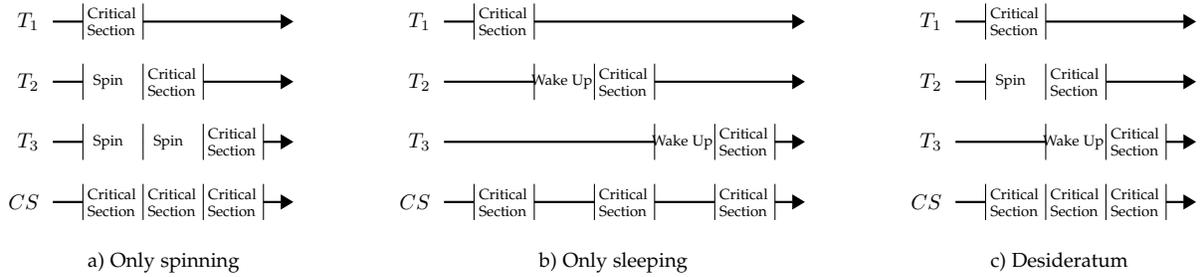
\begin{figure*}[t]
\centering
\begin{tikzpicture}

\begin{scope}[scale=0.8, every node/.append style={transform shape},xshift=-1cm]
\def \d {3}
\def \start {0.5}
\def \i {1}
\def \myfontsize {\scriptsize}
\def \len {4}

\draw[thick, -Triangle] (0,2) node[left=0.1]{$T_1$} -- +(\len,0);
\draw[lightgray!0!white, line width=4pt] (\start,2) -- +(\i,0) node[font=\myfontsize,black,right=-1.1,text width=30]{Critical Section};
\draw (\start+0*\i,2.\d) -- +(0,-2*0.\d);
\draw (\start+1*\i,2.\d) -- +(0,-2*0.\d);

\draw[thick, -Triangle] (0,1) node[left=0.1]{$T_2$} -- +(\len,0);
\draw[lightgray!0!white, line width=4pt] (\start+0*\i,1) -- +(\i,0) node[font=\myfontsize,black,left=0.15]{Spin};
\draw[lightgray!0!white, line width=4pt] (\start+1*\i,1) -- +(\i,0) node[font=\myfontsize,black,right=-1.1,text width=30]{Critical Section};
\draw (\start+0*\i,1.\d) -- +(0,-2*0.\d);
\draw (\start+1*\i,1.\d) -- +(0,-2*0.\d);
\draw (\start+2*\i,1.\d) -- +(0,-2*0.\d);

\draw[thick, -Triangle] (0,0) node[left=0.1]{$T_3$} -- +(\len,0);
\draw[lightgray!0!white, line width=4pt] (\start+0*\i,0) -- +(\i,0) node[font=\myfontsize,black,left=0.15]{Spin};
\draw[lightgray!0!white, line width=4pt] (\start+1*\i,0) -- +(\i,0) node[font=\myfontsize,black,left=0.15]{Spin};
\draw[lightgray!0!white, line width=4pt] (\start+2*\i,0) -- +(\i,0) node[font=\myfontsize,black,right=-1.1,text width=30]{Critical Section};
\draw (\start+0*\i,0.\d) -- +(0,-2*0.\d);
\draw (\start+1*\i,0.\d) -- +(0,-2*0.\d);
\draw (\start+2*\i,0.\d) -- +(0,-2*0.\d);
\draw (\start+3*\i,0.\d) -- +(0,-2*0.\d);

\draw[thick, -Triangle] (0,-1) node[left=0.1]{$CS$} -- +(\len,0);
\draw[lightgray!0!white, line width=4pt] (\start+0*\i,-1) -- +(\i,0) node[font=\myfontsize,black,right=-1.1,text width=30]{Critical Section};
\draw[lightgray!0!white, line width=4pt] (\start+1*\i,-1) -- +(\i,0) node[font=\myfontsize,black,right=-1.1,text width=30]{Critical Section};
\draw[lightgray!0!white, line width=4pt] (\start+2*\i,-1) -- +(\i,0) node[font=\myfontsize,black,right=-1.1,text width=30]{Critical Section};
\draw (\start+0*\i,-1+0.\d) -- +(0,-2*0.\d);
\draw (\start+1*\i,-1+0.\d) -- +(0,-2*0.\d);
\draw (\start+2*\i,-1+0.\d) -- +(0,-2*0.\d);
\draw (\start+3*\i,-1+0.\d) -- +(0,-2*0.\d);

\draw[lightgray!0!white, line width=4pt] (\start+1*\i,-2)  node[black,right=-1.1]{a) Only spinning};

\end{scope}

\begin{scope}[scale=0.8, every node/.append style={transform shape},xshift=5.5cm]
\def \d {3}
\def \start {0.5}
\def \i {1}
\def \myfontsize {\scriptsize}
\def \len {6}

\draw[thick, -Triangle] (0,2) node[left=0.1]{$T_1$} -- +(\len,0);
\draw[lightgray!0!white, line width=4pt] (\start,2) -- +(\i,0) node[font=\myfontsize,black,right=-1.1,text width=30]{Critical Section};
\draw (\start+0*\i,2.\d) -- +(0,-2*0.\d);
\draw (\start+1*\i,2.\d) -- +(0,-2*0.\d);

\draw[thick, -Triangle] (0,1) node[left=0.1]{$T_2$} -- +(\len,0);
\draw[lightgray!0!white, line width=4pt] (\start+1*\i,1) -- +(\i,0) node[font=\myfontsize,black,left=-0.15]{Wake Up};
\draw[lightgray!0!white, line width=4pt] (\start+2*\i,1) -- +(\i,0) node[font=\myfontsize,black,right=-1.1,text width=30]{Critical Section};
\draw (\start+1*\i,1.\d) -- +(0,-2*0.\d);
\draw (\start+2*\i,1.\d) -- +(0,-2*0.\d);
\draw (\start+3*\i,1.\d) -- +(0,-2*0.\d);

\draw[thick, -Triangle] (0,0) node[left=0.1]{$T_3$} -- +(\len,0);
\draw[lightgray!0!white, line width=4pt] (\start+3*\i,0) -- +(\i,0) node[font=\myfontsize,black,left=-0.15]{Wake Up};
\draw[lightgray!0!white, line width=4pt] (\start+4*\i,0) -- +(\i,0) node[font=\myfontsize,black,right=-1.1,text width=30]{Critical Section};
\draw (\start+3*\i,0.\d) -- +(0,-2*0.\d);
\draw (\start+4*\i,0.\d) -- +(0,-2*0.\d);
\draw (\start+5*\i,0.\d) -- +(0,-2*0.\d);

\draw[thick, -Triangle] (0,-1) node[left=0.1]{$CS$} -- +(\len,0);
\draw[lightgray!0!white, line width=4pt] (\start+0*\i,-1) -- +(\i,0) node[font=\myfontsize,black,right=-1.1,text width=30]{Critical Section};
\draw[lightgray!0!white, line width=4pt] (\start+2*\i,-1) -- +(\i,0) node[font=\myfontsize,black,right=-1.1,text width=30]{Critical Section};
\draw[lightgray!0!white, line width=4pt] (\start+4*\i,-1) -- +(\i,0) node[font=\myfontsize,black,right=-1.1,text width=30]{Critical Section};
\draw (\start+0*\i,-1+0.\d) -- +(0,-2*0.\d);
\draw (\start+1*\i,-1+0.\d) -- +(0,-2*0.\d);
\draw (\start+2*\i,-1+0.\d) -- +(0,-2*0.\d);
\draw (\start+3*\i,-1+0.\d) -- +(0,-2*0.\d);
\draw (\start+4*\i,-1+0.\d) -- +(0,-2*0.\d);
\draw (\start+5*\i,-1+0.\d) -- +(0,-2*0.\d);

\draw[lightgray!0!white, line width=4pt] (\start+2*\i,-2)  node[black,right=-1.1]{b) Only sleeping};

\end{scope}

\begin{scope}[scale=0.8, every node/.append style={transform shape},xshift=14cm]
\def \d {3}
\def \start {0.5}
\def \i {1}
\def \myfontsize {\scriptsize}
\def \len {4}

\draw[thick, -Triangle] (0,2) node[left=0.1]{$T_1$} -- +(\len,0);
\draw[lightgray!0!white, line width=4pt] (\start,2) -- +(\i,0) node[font=\myfontsize,black,right=-1.1,text width=30]{Critical Section};
\draw (\start+0*\i,2.\d) -- +(0,-2*0.\d);
\draw (\start+1*\i,2.\d) -- +(0,-2*0.\d);

\draw[thick, -Triangle] (0,1) node[left=0.1]{$T_2$} -- +(\len,0);
\draw[lightgray!0!white, line width=4pt] (\start+0*\i,1) -- +(\i,0) node[font=\myfontsize,black,left=0.15]{Spin};
\draw[lightgray!0!white, line width=4pt] (\start+1*\i,1) -- +(\i,0) node[font=\myfontsize,black,right=-1.1,text width=30]{Critical Section};
\draw (\start+0*\i,1.\d) -- +(0,-2*0.\d);
\draw (\start+1*\i,1.\d) -- +(0,-2*0.\d);
\draw (\start+2*\i,1.\d) -- +(0,-2*0.\d);

\draw[thick, -Triangle] (0,0) node[left=0.1]{$T_3$} -- +(\len,0);
%\draw[lightgray!0!white, line width=4pt] (\start+0*\i,0) -- +(\i,0) node[font=\myfontsize,black,left=0.1]{Sleep};
\draw[lightgray!0!white, line width=4pt] (\start+1*\i,0) -- +(\i,0) node[font=\myfontsize,black,left=-0.15]{Wake Up};
\draw[lightgray!0!white, line width=4pt] (\start+2*\i,0) -- +(\i,0) node[font=\myfontsize,black,right=-1.1,text width=30]{Critical Section};
%\draw (\start+0*\i,0.\d) -- +(0,-2*0.\d);
\draw (\start+1*\i,0.\d) -- +(0,-2*0.\d);
\draw (\start+2*\i,0.\d) -- +(0,-2*0.\d);
\draw (\start+3*\i,0.\d) -- +(0,-2*0.\d);

\draw[thick, -Triangle] (0,-1) node[left=0.1]{$CS$} -- +(\len,0);
\draw[lightgray!0!white, line width=4pt] (\start+0*\i,-1) -- +(\i,0) node[font=\myfontsize,black,right=-1.1,text width=30]{Critical Section};
\draw[lightgray!0!white, line width=4pt] (\start+1*\i,-1) -- +(\i,0) node[font=\myfontsize,black,right=-1.1,text width=30]{Critical Section};
\draw[lightgray!0!white, line width=4pt] (\start+2*\i,-1) -- +(\i,0) node[font=\myfontsize,black,right=-1.1,text width=30]{Critical Section};
\draw (\start+0*\i,-1+0.\d) -- +(0,-2*0.\d);
\draw (\start+1*\i,-1+0.\d) -- +(0,-2*0.\d);
\draw (\start+2*\i,-1+0.\d) -- +(0,-2*0.\d);
\draw (\start+3*\i,-1+0.\d) -- +(0,-2*0.\d);

\draw[lightgray!0!white, line width=4pt] (\start+1*\i,-2)  node[black,right=-1.1]{c) Desideratum};

\end{scope}

\end{tikzpicture}
%\vspace*{-0.3cm}
\caption{Different timelines related to different lock specifications}
%Timelines representing a synthetic scenario with three threads contending for acquire three different lock implementations. The %timeline at the bottom represent the projection of critical section execution on the real time axis.}
\label{fig:desiderata}
\end{figure*}

A lock implementation which copes with the issue of 
%tries to overcome the challenge of 
choosing at runtime between spinning and sleeping is the \emph{mutex} offered by the {\sf glibc} {\tt pthread} library\cite{glibc}.
This lock can work with two different behaviors: default and adaptive.
%denoted as {\tt PT-DEFAULT-MUTEX} and {\tt PT-ADAPTIVE-MUTEX} respectively.
%The goal of the default configuration is to make the uncontended case as fast as possible.
%In this setting, 
In the default configuration, a thread tries to acquire the lock by initially performing an atomic test-and-set operation.
If this operation fails,  
%the lock is acquired and the thread can execute the critical section, otherwise it 
the thread goes to sleep.
Conversely, the adaptive behavior is based on the idea of attempting to spin for a while before going to sleep. 
The limitation of this approach is that it does not offer any support for making a thread transiting from the sleep phase to the spin phase before the critical section is released. 
Hence, like for the case of pure sleep locks, the access to the critical section might be delayed because of latencies associated with OS-level awakening of a waiting thread upon mutex release. 
In other words, the adaptive mutex attempts to tackle the problem of reducing the waste of resources caused by excessive spin operations, but does not jointly copes with the optimization of the latency for accessing the critical section when sleeps occur, an issue that is instead tackled by our mutable locks.

\section{Mutable Locks}
\label{mutable}
%Before introducing the internals of our mutable lock, we want to explain the main idea behind them.
%For this reason, consider a 

Let us slide towards the description of our mutable locks
%algorithm 
through the help of an example scenario where 3 threads compete for the access to a critical section. For simplicity, but with no loss of generality, we consider the case where the critical section duration is equal to the time required by a thread to be awaken and CPU-rescheduled---if originally sleeping because of lock occupancy by another thread---and where each thread runs on a different CPU-core.
%, so that no
%hence no competition for CPU-core usage appears.
Figure \ref{fig:desiderata} shows different timelines resulting from the abovementioned workload running on different lock specifications.
%Each subfigure presents the timeline of each thread with its active states (executing critical sections, waking up and spinning) %and without considering sleeping phases and what happens after the critical section execution.
The timeline at the bottom shows the projection of each critical section on the real time axis.

Figure \ref{fig:desiderata}a) represents a possible execution resulting by adopting a spin lock (e.g. a test-and-test-and-set spin lock).
As we can see, threads that have loosen the challenge are always ready to participate to a new challenge.
This makes critical section executions immediately consecutive along the real time axis, requiring 3 slots for executing critical sections (CSes) and 3 slots for spinning.
It follows that the 50\% of the clock cycles dedicated to the execution of those three CSes are ``wasted'' in spin operations for ensuring minimum latency.

Figure \ref{fig:desiderata}b) shows the effects of always going to sleep if the lock is already taken by some other thread---this is the same strategy adopted by the pthread mutex in the default configuration.
During the release phase, each thread wakes only one thread at a time.
Consequently, we have to pay some awakening latency for accessing the critical section. 
In this scenario, 5 slots (each lasting the duration of the critical section) are required to complete 3 critical sections.
It follows that the overall throughput is 40\% worse than the one achieved by the spin lock-based approach, but 2 slots instead of 3 are wasted for awakening and CPU-reschedule operations.
%we are wasting the 40\% of the slots for awake and CPU-reschedule operations.
%---clearly the longer the lock duration %is, the lower clock-cycle waste will be in this setting.

Finally, Figure \ref{fig:desiderata}c) shows an optimized behavior with the same amount (2) of wasted slots as for the classical sleep-based approach, but where the latency for accessing the critical section is the same as the one of the spin-lock approach. 
%This is possible because 
In this scenario the lock is able to decide which thread has to go to sleep and which one to spin during the challenge for acquiring it, or to transit out of the sleep state before attempting to reacquire the lock.
%the critical secition is released by some other thread.
In particular, it makes the latency of awakening a thread ($T_3$) be masked by the critical section execution of the spinning thread ($T_2$). This optimized behavior is the target of our mutable lock algorithm, which encapsulates the ability to mix spin and sleep phases in an optimized manner, with the inclusion of sleep to spin transitions.
%In a more general case, we need $N$ critical sections to mask the latecy a single wake up.
%This can be achieved by making up to $N$ threads spin.

%\subsubsection{/*SPINNING WINDOW SPECIFICATION*/}

%In order to formalize the idea of mixing spin and sleep phases, and introducing the anticipated transition frm sleep to spin, %our mutable locks are based on the concept of

\begin{figure}[t]
\centering
\begin{tikzpicture}[scale=0.7]

\def \start {2}
\def \m{0.6}
\def \myfontsize {\scriptsize}
\def \len {11}

\draw[thick] (0,\start+\m) node[left=0.1]{} -- +(\len,0);
\draw[thick] (0,\start) node[left=0.1]{} -- +(\len,0);

\foreach \i in {-1,0,4,7}{
\draw[thick] (\i+1,\start) -- +(0,\m);
}

\foreach \i [evaluate=\i as \j using {int(\i+1)}] in {-1,...,9}{
\draw ( \i+1 ,\start) -- +(0,\m)node[font=\myfontsize,above]{\j};
}

\foreach \i in {0,...,7}{
\node at (\i+0.5,\start + \m/2) {X};
}
\draw[lightgray!0!white, line width=4pt] (10,\start + \m/2) -- +(1.5,0) node[font=\myfontsize,black,right=-1.0]{........};

\draw[-Triangle] (0.5,\start+\m*2) node[font=\myfontsize,above, align=center]{Thread\\ in critical section} --  (0.5,\start+\m*4/3) node{};
\draw[-Triangle] (5,\start+\m*2.5) node[font=\myfontsize,above]{Spinning Window Size} --  (5,\start+\m+\m*0.8) node{};

\draw[decorate,decoration={brace,amplitude=10pt,mirror}](1,1.8) -- +(4,0)  node[font=\myfontsize,midway,below=0.5]{Spinning Window};
\draw[-Triangle] (8,1.6) node[font=\myfontsize,below]{Thread Count} --  (8,1.9) node{};
\end{tikzpicture}
%\vspace*{-0.8cm}
\caption{Logical representation of a lock with a spinning window}
\label{fig:mlock}
\end{figure}
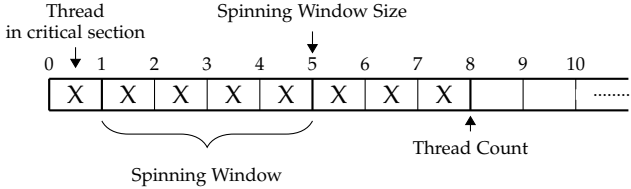

\subsection{The Notion of Spinning Window}
A baseline concept the mutable lock relies on is the \emph{spinning window} (SW).
%Among the treads contending for the critical section, 
SW allows identifying a set of threads allowed to spin---among those contending for the critical section---
while the others (if any) need to undergo a sleep phase. The maximum cardinality of this set is bounded by an integer value called \emph{spinning window size} (SWS).
A logical representation of the effects of using SW is shown in Figure \ref{fig:mlock}.
Here we have an array where each slot is occupied by a waiting thread except for the first one (with index equal to 0) that represents the thread holding the lock.
The next SWS cells form the SW and are occupied by threads allowed to spin, while others are outside the SW and are sleeping.
A new arriving thread $T$ takes the first  available slot and, according to its index $i$, it choses if it has to spin or to sleep.
In particular:
{\begin{gather*}
\small
\begin{cases}
i = 0 & \text{$T$ grubs the lock}\\
i \in [1,SWS] & \text{$T$ goes to spin}\\
i \in (SWS,+\infty) & \text{$T$ goes to sleep}\\
\end{cases}
\end{gather*}}
The lock release operation consists in making one random spinning thread access the critical section and one random sleeping thread wake up and occupy the just freed slot of the SW. The latter is, essentially, the sleep to spin transition we include in our mutable lock logic.
Then, the array cell that is left empty outside the spinning window by the woke up thread will be occupied by shifting the threads associated with larger indexes. 
This complies with a specification that does not ensure a FIFO policy while serving threads---which can be instead obtained by left-shifting all threads in the array exactly by one position.
It follows that this approach allows to control the exact number of threads allowed to spin---including those transiting from the sleep to spin state---by manipulating the value of SWS.

Since we are interested in pursuing two goals, maximizing performance and reducing the waste of clock cycles caused by spin operations, the SWS should adapt to the workload peculiarities and changes---e.g. to the duration of the critical section and the actual incidence of conflicts in its access. The larger SWS, the lower the access latency---although if too many threads spin, we get CPU-interference on the one running the critical section--but, at the same time, more computational power is wasted due to spinning cycles.
Conversely, a lower value of SWS tends to reduce clock cycles usage, but increases the probability that threads experience late wake ups, stretching the critical section access latency. 
Overall, a suited value for SWS is the one ensuring that the number of spinning threads is low and the latency of awakening threads is masked by critical section executions by other threads.

%In a real implementation, d
Dynamically adapting the value of SWS at runtime is not a trivial task since the newly chosen value must be correctly reflected onto the actual state of the threads (sleeping or spinning). 
%a new value cloud 
%lead the mutable lock state to be no longer compliant with the actual ststes if the threads using it. 
%might require some additional caution, since the new state of the lock could be not coherent with actual state of each thread accessing the lock.
%For example, consider the scenario in Figure \ref{fig:mlock}.
More in detail, increasing the value of SWS with no other action could make one or more threads to be considered as falling within SW (thus spinning) even if they are currently sleeping (case 1).
Consequently, no one will ever try to wake them up and they will sleep unboundedly unless the SWS is eventually restored to the original value.
%In order to avoid this scenario is enough waking up those threads contextually with the SW resize. 
On the other hand, reducing the value of SWS, which might make some spinning thread be outside SW (case 2), does not hamper progress.
However, it makes a number of threads larger than SWS spin for an unknown period of time, diverging from the desidered behavior.
%To prevent this, we have to keep track of those spinning threads during a resize and give them an higher priority while moving %threads within the SW during a release phase, e.g. by waking none up for some iterations.

Case 1 and 2 can occur only under specific conditions.
Let $\Delta$ be the variation of the value of SWS to be applied at runtime, $sws$ the value of SWS before the variation is applied, and $thc$ (thread count) be the number of threads waiting to access the lock.
%and $sws$ be the actual SWS value once applied the change $\Delta$.
Case 1 can occur if and only if
%\begin{equation}\tag{C1}
$\Delta > 0~~ \wedge ~~ thc > sws$ (C1) 
%\label{eq:case1}
%\end{equation}
%\noindent
while case 2 can occur if and only if 
%\begin{equation}\tag{C2}
$\Delta < 0~~ \wedge ~~thc > sws + \Delta$ (C2).
%\label{eq:case2}
%\end{equation}

%Our mutable lock algorithm implements an integrated logic, embedded within the lock acquisition and release API, 
The SW specification tackle case 1 by waking up a number of sleeping threads equal to $\min(|\Delta|, thc - sws)$ instead of 1 as in normal release operations. 
Case 2 is tackled by assigning to a number of spinning threads higher priority in the access to SW, with respect to threads that transit from the sleep to the spin state. 
This can be obtained by simply waking no thread up for a number of release phases equal to $\min(|\Delta|, thc - (sws + \Delta))$. 
%Clearly, the change of the state of threads (or assigment of priorities for the access to SW) on the basis of the current value of SWS and its vatiations 

This \emph{mutable lock} algorithm, described in Section \ref{mlalgo}, is de-facto a 
%the manipulation of the value of SWS and the consequent
%needs to be carried out within a proper algorithm, which is essentially a 
new thread synchronization algorithm grounded on the notion of locking primitive.
%embedded within the acqusition and release operations of the mutable lock. 
%This algorithm 
%which inmplements the mutable lock logic, as 
%will be described in Section \ref{mlalgo}.
At the same time, the change of SWS, in order to adapt its value to the workload, needs to be actuated via some other algorithm, which implements a kind of {\em oracle} for optimizing the runtime dynamics under mutable lock-based synchronization. 
%the mutable lock algorithm. 
%One of these oracles is provided in Section \ref{oracle}. However, we note that 
Clearly, different oracles for adapting the SWS value at runtime could be devised, and we provide one of them in Section \ref{mlalgo}. In any case, the mutable lock algorithm is independent of the actually selected SWS adaptation oracle. This opens to the possibility of studying further variations of thread synchronization dynamics built around the notion of mutable lock. 

%we present works correctly independently by the fact that one or another oracle tells that some specific chage needs %to be applied to the value of SWS. 
%as complements to mutable lock algorithm, which is therefore able to handle whathever chage of SWS is imposed by the %seleted oracle. 
%This logic is essentially a synchronizaiton algorithm---to be presented in the next section---that manipulates the threads %states and the variables representing the state of the mutable lock in a combined manner.

\subsection{The Mutable Lock Algorithm}
\label{mlalgo}

%In this section, we propose an implementation of a lock, called \emph{mutlock}, implementing the SW mechanism. 
A mutable lock is a spin lock, denoted as \slock, plus other five variables:
%
%\begin{itemize}
%\vspace*{-0.3cm}\item \sws, which stores the current \emph{spinning window size};
%%, namely the maximum cardinality of the set containing threads allowed to spin, thus, excluding the one holding the lock;
%\item \thc, which keeps the \emph{thread count}, i.e. the number of threads currently waiting for accessing the lock;
%%needing to acquire the lock (this counter includes the thead that already keeps the lock, if any);
%%accessing the lock, whose default value is -1;
%\item \wuc, which keeps the \emph{wake-up count}, i.e.  number of threads to be woken up during a mutable lock release phase;
%\item \futex, which is a blocking synchronization object used to access the sleep/wake-up API of the underlying Operating System;
%\item {\tt max\_sws} is the maximum SWS set to the number of cores.
%\end{itemize}
%
%\vspace*{-0.3cm}
\sws, which stores the current \emph{spinning window size};
\thc, which keeps the \emph{thread count}, i.e. the number of threads currently waiting for accessing the lock plus one (that holds the lock);
\wuc, which keeps the \emph{wake-up count}, i.e.  number of threads to be woken up during a mutable lock release phase;
\futex, which is a blocking synchronization object used to access the sleep/wake-up API of the underlying Operating System;
{\tt max} is the maximum SWS set to the number of cores.

In our design, \sws\ and \thc\ are 32 bits long and are stored in a unique 64-bits word, denoted as \lstate\ (lock state), such that $\lstate=\langle \sws,\thc \rangle$.
This arrangement allows threads to update one field, get its old value and retrieve the actual value of the other field in once by using an atomic {\tt Fetch\&Add} (FAD) machine instruction, commonly supported by off-the-shelf processors. 
%In order to ensure that multiple SW resizes do not occur concurrently, we allow only the thread holding the lock to update \sws\ with a \FAD\ on the most significat 32 bits of \lstate.
%remember that threads outside the critical section might update and read \thc, namely the lowest significant 32 bits of \lstate.

\begin{algorithm*}[t]
\algosizefinal
\caption{Mutable Lock Operations}\label{algo}
\begin{minipage}[t]{0.48\linewidth}
  \algrenewcommand{\alglinenumber}[1]{\algprintlinenumber{ \makebox[20pt][r]{\algosizefinal A#1:}}}
  \begin{algorithmic}[1]
    \Procedure{Acquire}{mutlock m}
      \State $\Delta$ $\leftarrow$ $0$
      \State $slept$ $\leftarrow$ $false$
      \State $lstate^-$ $\leftarrow$ {\tt FAD}(m.\lstate, +1) \label{line:INC}
      \State $thc^-$ $\leftarrow$ $lstate^-.\thc$
      \State $sws$ $\leftarrow$ $lstate^-.\sws$
      \If {$thc^-$ $\geq$ $sws$ \label{line:sleepCond}}
        \State $slept$ $\leftarrow$ $true$
        \State $m.\futex$.sleep() \label{line:goSleep}
      \EndIf
      \State $spun$ $\leftarrow$ $m.\slock$.lock() \label{line:goSpin}
      \State $\Delta$ $\leftarrow$ EvalSWS($spun$, $slept$, $m$)
      \If {$sws$ $\not=$ $m.\lstate.\sws$}
        \State {\bf return}
      \EndIf
      \State $\Delta$ $\leftarrow$ $sws+\Delta$ $<$ $1$ ? $sws-1$ : $\Delta$
      \State $\Delta$ $\leftarrow$  $sws+\Delta$ $>$ $m$.{\tt max} ? $m$.{\tt max}$-sws$ : $\Delta$
      \If {$\Delta \not= 0$}
        \State $tmp$ $\leftarrow$ ($\Delta$ {\tt <<} 32);   
        \State $lstate^-$ $\leftarrow$ {\tt FAD}(m.\lstate, $tmp$) \label{line:updateSWS}
        \State $thc$ $\leftarrow$ $lstate^-.\thc$
        \State $sws^-$ $\leftarrow$ $lstate^-.\sws$
        \State $tmp$ $\leftarrow$ $+\infty$
        \State $sign$ $\leftarrow$ $\Delta/|\Delta|$      
        \If {$sign$ $<$ $0$ $\wedge$ $\thc$ $>$ $\sws^+$}
          \State $tmp$ $\leftarrow$ $\thc-\sws^+$
        \ElsIf {$sign$ $>$ $0$ $\wedge$ $\thc$ $>$ $\sws^-$}
          \State $tmp$ $\leftarrow$ $\thc-\sws^-$
        \Else
          \State $tmp$ $\leftarrow$ $0$ \label{line:tmp0}
        \EndIf
        \State $tmp$ $\leftarrow$ $sign \cdot \min(|\Delta|,tmp)$
        \State $m.\wuc$ $\leftarrow$ $m.\wuc$ + $tmp$ \label{line:updateWUC}
      \EndIf
    \EndProcedure
  \end{algorithmic}
\end{minipage}
\hfill
\begin{minipage}[t]{0.48\linewidth}
  \algrenewcommand{\alglinenumber}[1]{\algprintlinenumber{ \makebox[20pt][r]{\algosizefinal R#1:}}}
  \begin{algorithmic}[1]
    \Procedure{Release}{mutlock m}
    %\Return void %<void *, int>
      \If{$m.\wuc \geq 0$}
        \State $R_{wuc}$ $\leftarrow$ $m.\wuc$ \label{line:wucP}
        \State $m.\wuc$ $\leftarrow$ $0$
      \Else
        \State $R_{wuc}$ $\leftarrow$ $-1$ \label{line:wucN}
        \State $m.\wuc$ $\leftarrow$ $m.\wuc+1$
      \EndIf
      \State $lstate^-$ $\leftarrow$ {\tt FAD}(m.\lstate, -1) \label{line:dec}
      \State $m.\slock$.unlock() \label{line:release}
      \If {$R_{wuc}$ $<$ $0$}
        \State {\bf return}
      \EndIf
      \State $thc^-$ $\leftarrow$ $lstate^-.\thc$
      \State $sws$ $\leftarrow$ $lstate^-.\sws$
      \If {$thc^-$ $>$ $sws$ \label{line:wucInc}}
        \State $R_{wuc}$ $\leftarrow$ $R_{wuc}+1$
      \EndIf
      \While {$wuc>0$}
        \State $cnt$ $\leftarrow$ $m.\futex$.wake\_up($R_{wuc}$) \label{line:wakeup}
        \State $R_{wuc}$ $\leftarrow$ $R_{wuc}-cnt$
      \EndWhile
    \EndProcedure
  \end{algorithmic}
  \vspace{0.25cm}
  \hrule
  \vspace{0.25cm}
   \algrenewcommand{\alglinenumber}[1]{\algprintlinenumber{ \makebox[20pt][r]{\algosizefinal E#1:}}}
  \begin{algorithmic}[1]
    \Procedure{EvalSWS(}{}{bool $spun$, bool $slept$, mutlock $m$)}
      \State $m$.{\tt cnt} $\leftarrow$ $m$.{\tt cnt} $+$ $1$ 
      \State $\Delta$ $\leftarrow$ $0$
      \If {$slept$ $\wedge$ $\neg spun$}
        \State $\Delta$ $\leftarrow$ $m$.\sws
        \State $m$.{\tt cnt} $\leftarrow$ $0$ 
      \ElsIf {$m$.{\tt cnt} $=$ $K$}
        \State $\Delta$ $\leftarrow$ $-1$
        \State $m$.{\tt cnt} $\leftarrow$ $0$ 
      \EndIf
      \State \Return {$\Delta$} 
    \EndProcedure
  \end{algorithmic}
\end{minipage}
\label{alg:mutlock}
\end{algorithm*}

The operations used to acquire or release the mutable lock are shown in Algorithm \ref{alg:mutlock}.
%, but, before introducing them, we give some notations that should make the explanation more readable.
Let $x$ be an atomic register (a variable) supporting atomic FAD operations, in our notation $x^-$ and $x^+$ are the values of $x$ respectively before and after a FAD execution on it. 
During an acquire phase, a thread $T$ 
%checks if it can acquire the lock or there is space in the spinning window by 
increasing \thc\ via FAD (line A\ref{line:INC}) and checks whether there is space in SW.
If the condition $\thc^- \geq \sws$  holds (line A\ref{line:sleepCond})---no room is available in SW---it goes to sleep on \futex\ (line A\ref{line:goSleep}). Otherwise 
%($\thc^- < \sws$) 
it invokes the acquire API of \slock\ (line A\ref{line:goSpin}).

As soon as a thread owns the \slock, it determines if \sws\ should be updated by invoking {\sc EvalSWS}. This is the function implementing the SWS adaptation oracle. It returns the signed variation $\Delta$ to be applied to \sws. 
This update is performed via FAD on the most 32 significant bits of the \lstate\ field (line A\ref{line:updateSWS}).
Based on the values of $\Delta$, \thc, $\sws^-$ and $\sws^+$, we know that some countermeasure has to be taken in order to ensure progress of each thread and that the number of spinning threads will be eventually bounded  by \sws.
In particular, if condition C1 occurs, we set the variable \wuc\ to the number of additional threads to be woken up.
Conversely, when condition C2 holds, \wuc\ will be set to the number of threads in a spinning state which are outside the SW multiplied by -1.
Finally, if none of the above conditions holds, no countermeasure is needed at all (line A\ref{line:tmp0}).
At this point, the computed value will be simply added to \wuc\ (line A\ref{line:updateWUC}),  and the lock acquire phase is completed.
%and the CS can be executed.

%The release operation starts by checking the value of \wuc.

Upon a lock release operation,  
if $\wuc\geq0$ holds (line R\ref{line:wucN}), its value is copied into a local variable $R_{wuc}$ and then is set to 0, otherwise (line R\ref{line:wucP}) it is incremented by 1 and $R_{wuc}$ is set to -1.
Now, the \thc\ can be decremented by 1 via FAD and the \slock\ release API 
%executed 
allows another thread to get the lock %access the CS 
(lines R\ref{line:dec} and R\ref{line:release}).
In order to complete the release operation, we have to ensure that the number of non-sleeping threads is compliant with the current value of \sws.
Thus, the releasing thread first check if $R_{wuc}$ is lower than 0.
In this case, it can simply return since a previous reduction of \sws\ has made some thread spinning outside the SW and, consequently, no additional wake up is required.
This is because \sws\ updates and decrements of \thc\ are performed in mutual exclusion (via FAD) and it is ensured that more than \sws\ threads are spinning. 
If $R_{wuc}$ is greater than or equal to 0, we need to check if an additional thread should be awakened in order to keep a number of spinning threads equal to the current value of \sws.
In this case (line R\ref{line:wucInc}), $R_{wuc}$ is incremented by 1.
Finally, the thread can awake $R_{wuc}$ threads by relying on the \futex\ API (line R\ref{line:wakeup}).
Thanks to these algorithms, shared variables ($thc,sws,wuc$) used to keep the state of the lock are updated consistently without resorting to additional locks that could lead to other challenges such as choosing the proper lock implementation for protecting them.

%Such an operation is performed in a retry loop since the increasing of \thc\ during an acquire phase and going to sleep are not a single atomic step, and a thread might have to wait that threads are effectively sleeping.

%As a last note, if both spin and sleep object guarantee are starvation-free (e.g. FIFO), our mutlock should guarantees %starvation-freedom, a progress condition ensured, at the best of our knowledge, by none of any locks combining both %sleep and spin waiting phases. 

%\subsection{An Oracle for Setting SWS}
%\label{oracle}

%\begin{wrapfigure}{tl}{0.5\textwidth}
%\vspace{-1.0cm}
%\begin{minipage}[t]{\linewidth}
%\begin{algorithm}[H]
%\algosizefinal
%\caption{SWS resize oracle}\label{eval}
%  \algrenewcommand{\alglinenumber}[1]{\algprintlinenumber{ \makebox[20pt][r]{\algosizefinal E#1:}}}
%  \begin{algorithmic}[1]
%    \Procedure{EvalSWS(}{}{bool $spun$, \\ \hfill bool $slept$, mutlock $m$)}
%      \State $m$.{\tt cnt} $\leftarrow$ $m$.{\tt cnt} $+$ $1$ 
%      \State $\Delta$ $\leftarrow$ $0$
%      \If {$slept$ $\wedge$ $\neg spun$}
%        \State $\Delta$ $\leftarrow$ $m$.\sws
%        \State $m$.{\tt cnt} $\leftarrow$ $0$ 
%      \ElsIf {$m$.{\tt cnt} $=$ $K$}
%        \State $\Delta$ $\leftarrow$ $-1$
%        \State $m$.{\tt cnt} $\leftarrow$ $0$ 
%      \EndIf
%      \State \Return {$\Delta$} 
%    \EndProcedure
%  \end{algorithmic}
%\end{algorithm}
%\end{minipage}
%\end{wrapfigure}
%\vspace{-0.5cm}

%Our mutlock faces this challenge by applying a simple resize policy (shown in Algorithm \ref{eval}) of the spinning %%window based on a simple observation.
The oracle (shown in the routine EvalSWS of Algorithm \ref{alg:mutlock}) we present for dynamically varying the SWS is based on the following policy. If a thread $A$ wakes up and there are no spinning threads, it means that SWS is not larger enough to mask wake-up latency.
In fact, we know that, when $A$ arrived to the lock, there were other threads in active wait ($thc \geq sws^-$) and those threads have consumed SWS critical sections.
In this case, we double SWS.
Conversely, if such an event does not occur for $K$ consecutive critical-section executions, the oracle tries to decrease SWS by 1.
This should allow us to keep the SWS below the minimum value required to mask the wake up latency in $1/(K+1)$ cases. 
Clearly, choosing the proper value for $K$ depends on characteristics specific of the underlying hardware/software stack and its trade-off between the impact of hardware contention and latencies for waking threads up.
However, finding the optimal oracle is beyond the scope of this work, which is focused on providing a new technique for combining spinning and sleeping waiting phases that, at the best of our knowledge, explores the usage of a state transition (sleep to spin) never adopted by previous approaches.

\section{Experimental Study}
\label{data}

%\includepdfmerge[nup=2x2, landscape] {
%plots/GenuineIntel-out-batch_res_0_3.7_0_3.7/csl-0-csu-3.7-ncsl-0-ncsu-3.7-TH.pdf,
%plots/GenuineIntel-out-batch_res_0_3.7_0_3.7/csl-0-csu-3.7-ncsl-0-ncsu-3.7-DIFF.pdf,
%plots/GenuineIntel-out-batch_res_0_366_0_3.7/csl-0-csu-366-ncsl-0-ncsu-3.7-TH.pdf,
%plots/GenuineIntel-out-batch_res_0_366_0_3.7/csl-0-csu-366-ncsl-0-ncsu-3.7-DIFF.pdf
%}

%\includepdfmerge[nup=2x2, landscape] {
%plots/GenuineIntel-out-batch_res_0_3.7_0_366/csl-0-csu-3.7-ncsl-0-ncsu-366-TH.pdf,
%plots/GenuineIntel-out-batch_res_0_3.7_0_366/csl-0-csu-3.7-ncsl-0-ncsu-366-DIFF.pdf,
%plots/GenuineIntel-out-batch_res_0_366_0_366/csl-0-csu-366-ncsl-0-ncsu-366-TH.pdf,
%plots/GenuineIntel-out-batch_res_0_366_0_366/csl-0-csu-366-ncsl-0-ncsu-366-DIFF.pdf
%}

\begin{figure*}[!ht]
\centering
\subfigure[]{      
\includegraphics[width=0.31\linewidth, clip, trim = 4 3 14 17]{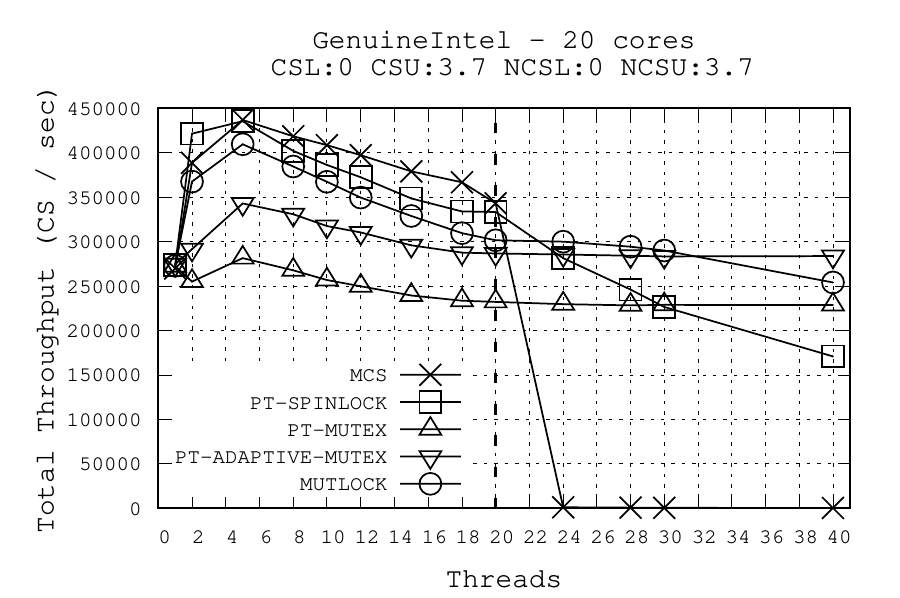} 
\label{fig:mltest1}
}
\subfigure[]{
\includegraphics[width=0.31\linewidth, clip, trim = 4 3 14 17]{plots/GenuineIntel-out-batch_res_0_3.7_0_3.7/{csl-0-csu-3.7-ncsl-0-ncsu-3.7-DIFF}.pdf}                                           
\label{fig:mltest2}
}
\subfigure[]{
\includegraphics[width=0.31\linewidth, clip, trim = 4 3 14 17]{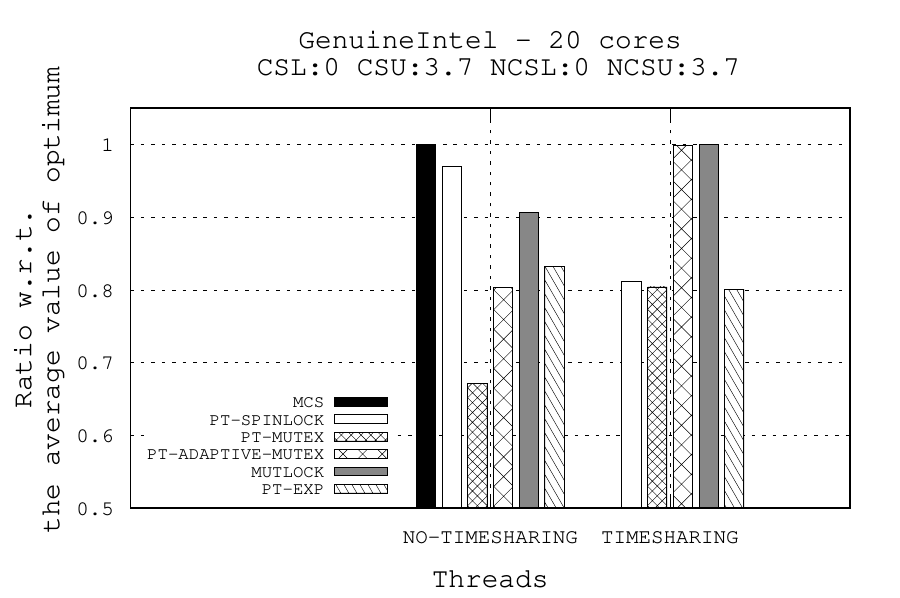}                                           
\label{fig:mltest10}
}
\\
\subfigure[]{      
\includegraphics[width=0.31\linewidth, clip, trim = 4 3 14 17]{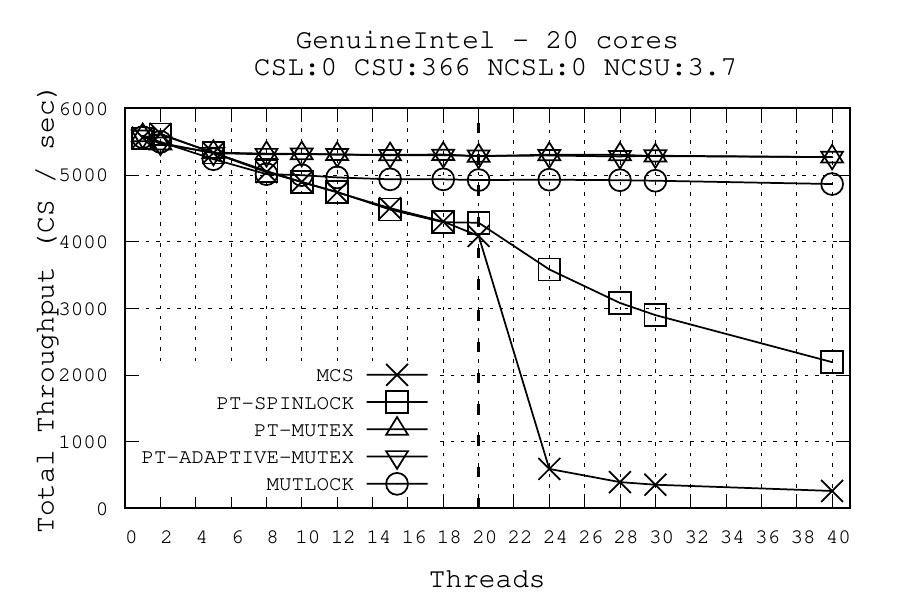}   
\label{fig:mltest3}
}
\subfigure[]{
\includegraphics[width=0.31\linewidth, clip, trim = 4 3 14 17]{plots/GenuineIntel-out-batch_res_0_366_0_3.7/{csl-0-csu-366-ncsl-0-ncsu-3.7-DIFF}.pdf}                                           
\label{fig:mltest4}
}
\subfigure[]{
\includegraphics[width=0.31\linewidth, clip, trim = 4 3 14 17]{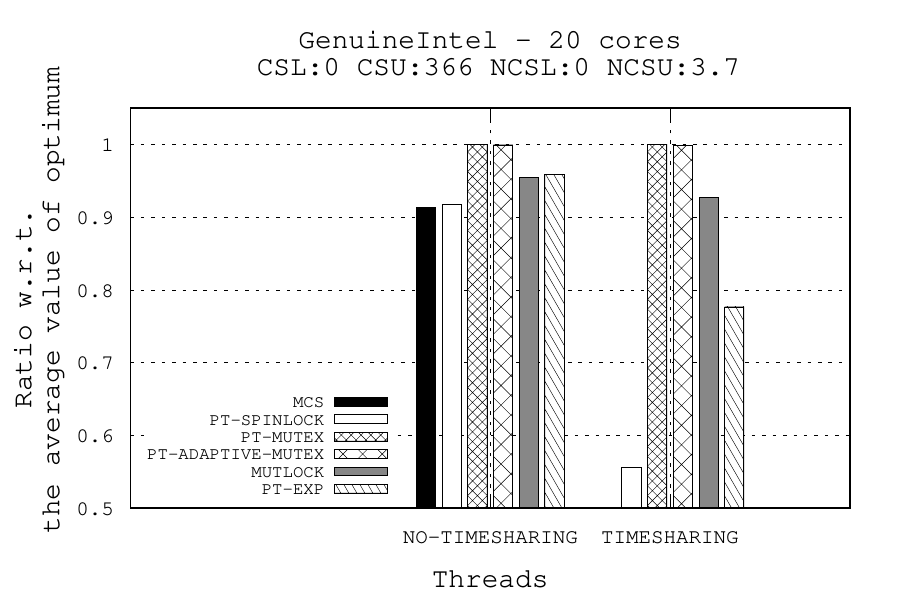}                                           
\label{fig:mltest11}
}
\\
\subfigure[]{      
\includegraphics[width=0.31\linewidth, clip, trim = 4 3 14 17]{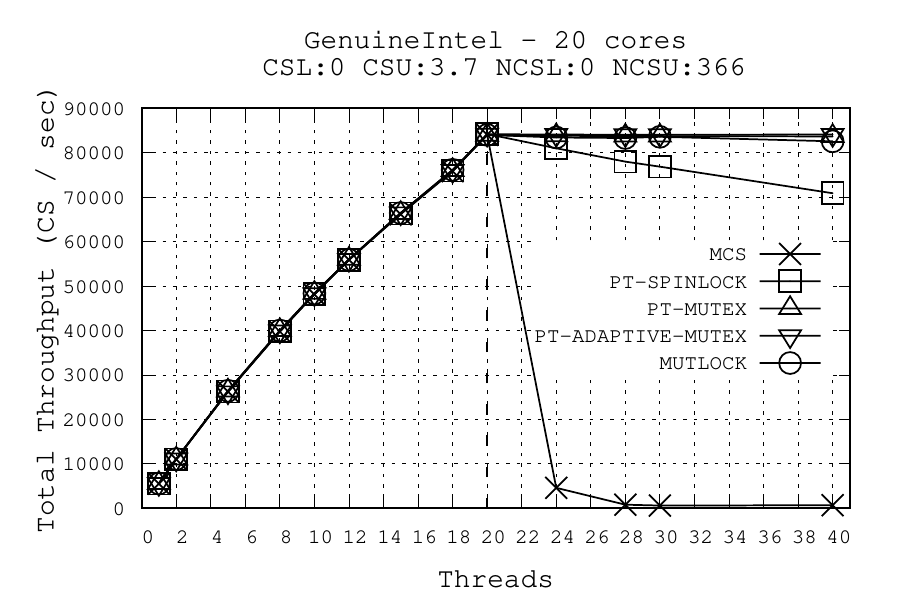}   
\label{fig:mltest5}
}
\subfigure[]{
\includegraphics[width=0.31\linewidth, clip, trim = 4 3 14 17]{plots/GenuineIntel-out-batch_res_0_3.7_0_366/{csl-0-csu-3.7-ncsl-0-ncsu-366-DIFF}.pdf}                                           
\label{fig:mltest6}
}
\subfigure[]{
\includegraphics[width=0.31\linewidth, clip, trim = 4 3 14 17]{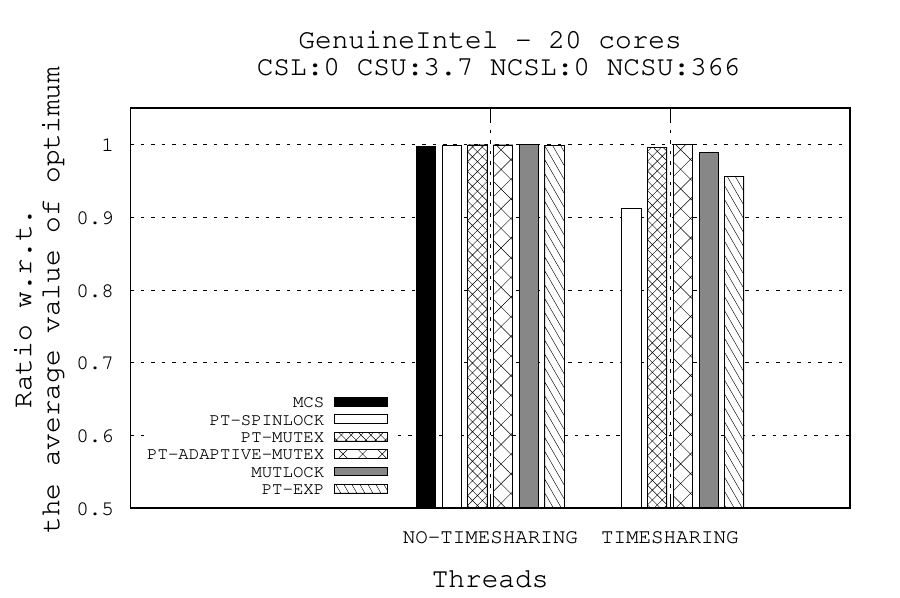}                                           
\label{fig:mltest12}
}
\\
\subfigure[]{      
\includegraphics[width=0.31\linewidth, clip, trim = 4 3 14 17]{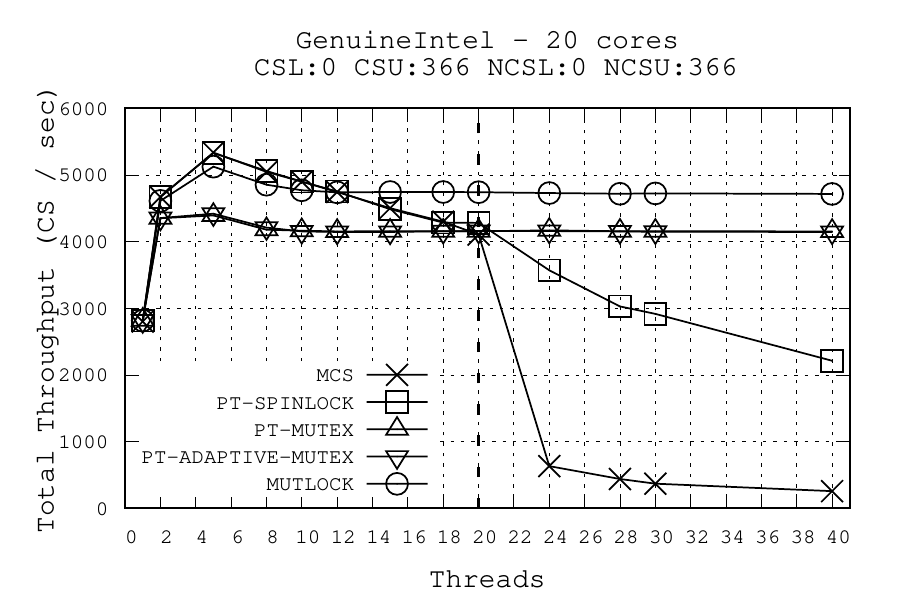} \vspace{0cm}  
\label{fig:mltest7}
}
\subfigure[]{
\includegraphics[width=0.31\linewidth, clip, trim = 4 3 14 17]{plots/GenuineIntel-out-batch_res_0_366_0_366/{csl-0-csu-366-ncsl-0-ncsu-366-DIFF}.pdf}                                           
\label{fig:mltest8}
}
\subfigure[]{
\includegraphics[width=0.31\linewidth, clip, trim = 4 3 14 17]{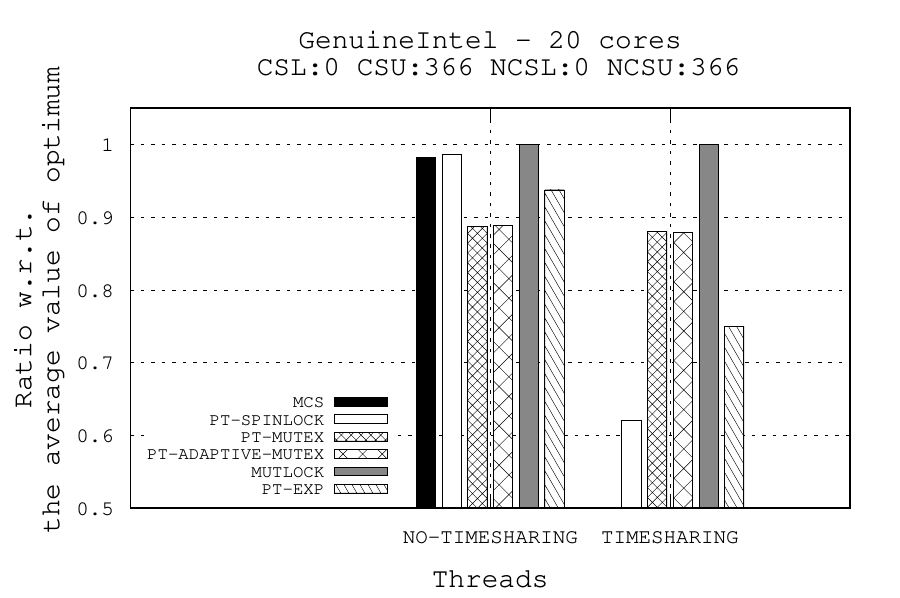}                                           
\label{fig:mltest13}
}
\caption{Tests consist in repeatedly executing critical and non-critical sections, whose lengths are uniformly distributed in the interval $[CSL, CSU]$ and $[NCSL, NCSU]$ respectively. Left charts show throughput (the higher the better), while ones in the middle column show CPU time spent in synchronization (the lower the better). Finally, charts on the right gives the ratio between the average throughput of a given lock and the average of the optimum obtained getting the lock with maximum throughtput for each thread count (the higher the better)---PT-EXP refers to the mean of PT-SPINLOCK and PT-MUTEX values.}
%\vspace*{-0.2cm}
%\caption{Results with the synthetic benchmark.}
\label{fig:expeval}
\end{figure*}

\begin{figure*}[!t]
\centering
\includegraphics[width=0.35\linewidth]{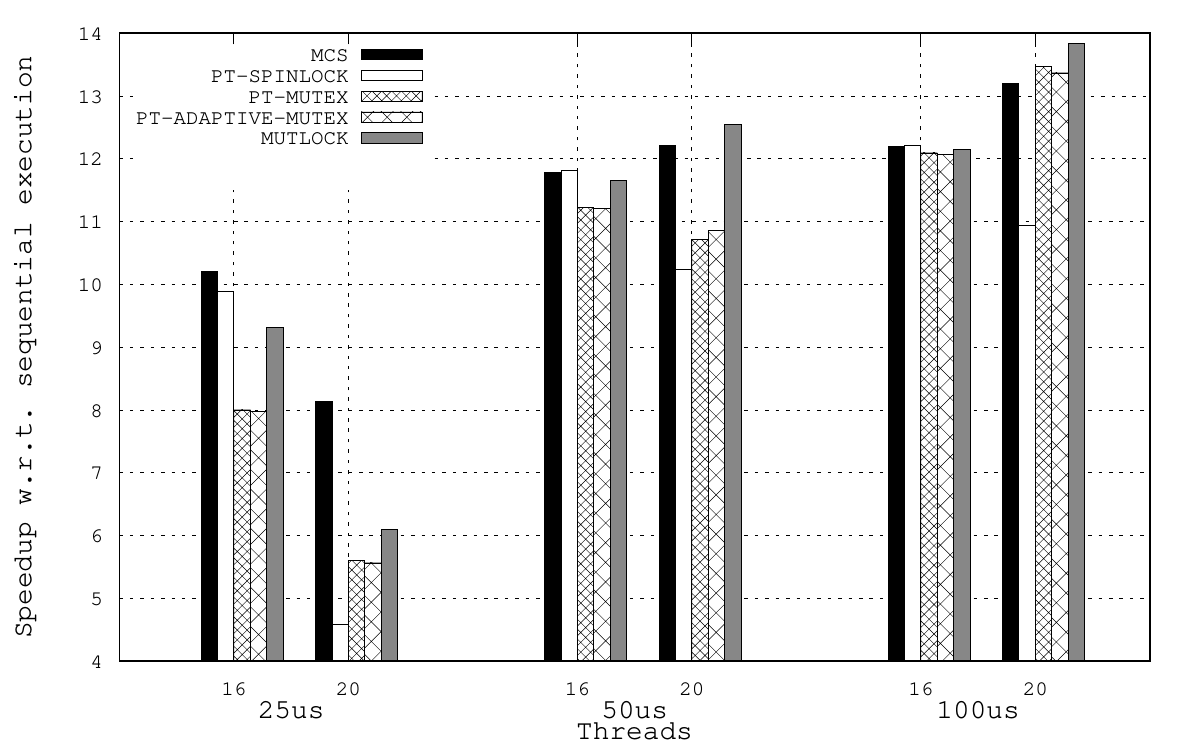}
\hspace{1cm}
\includegraphics[width=0.35\linewidth]{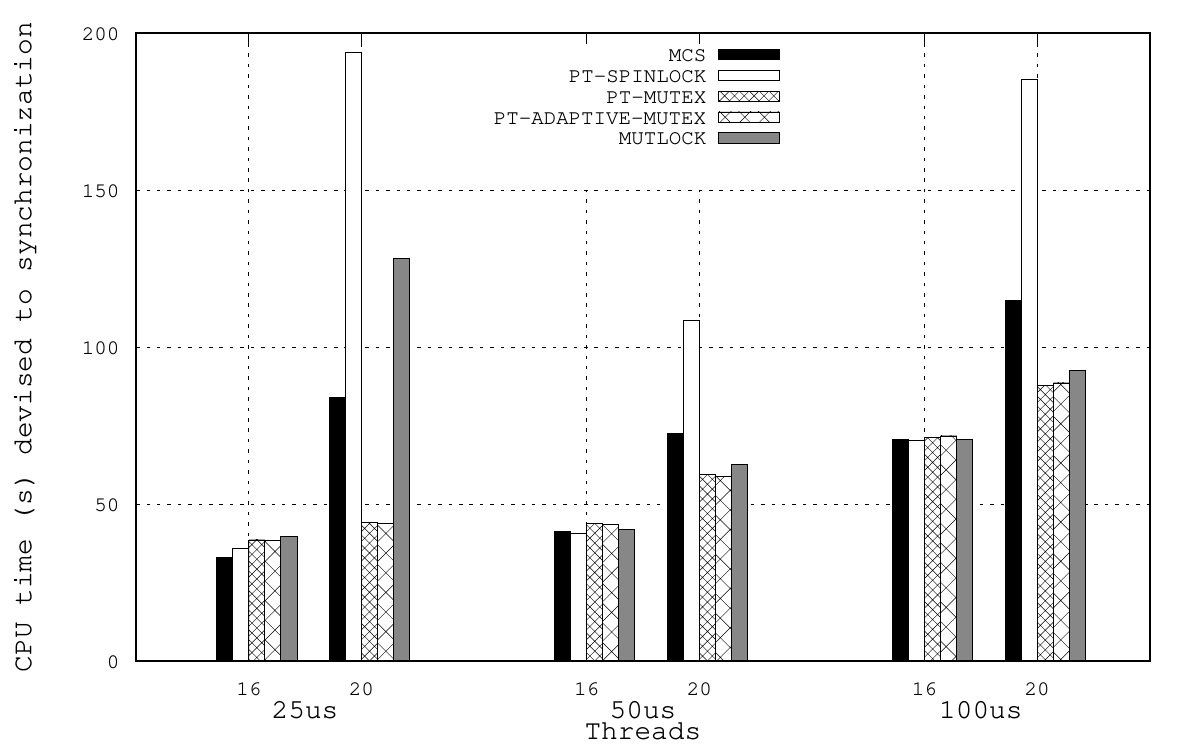}
\caption{Results for PHOLD run on top of a share-everything PDES platform.}
\label{fig:pdes}
\end{figure*}

We have evaluated our mutable lock (denoted as MUTLOCK) against the pthread spin lock (PT-SPINLOCK), an implementation of a queue lock (MCS) and the pthread mutex in both default and adaptive configuration (PT-MUTEX and PT-ADAPTIVE-MUTEX).
MUTLOCK adopts a classical test-and-test-and-set spin lock as \slock\ and a semaphore as sleeping object.
The parameter $K$ of the oracle has been set to 10 in order to keep the probability of paying the latency for waking threads up below the 10\%, making the mutable lock biased in avoiding late wake ups instead of reducing hardware contetion.
Each implementation has been evaluated by resorting to one synthetic benchmark and one real-world application.
The first one, called \emph{lockbench}\footnote{Available at \url{https://github.com/HPDCS}}, makes a given number of threads repeatedly access a critical section and then execute a non-critical section, whose lengths $CS$ and $NCS$ are uniformly distributed within a given interval equal to $[CSL,CSU)$ and $[NCSL,NCSU)$ respectively.
Our performance metric is the throughput intended as number of executed critical sections per unit time, while we have adopted the CPU time spent in synchronization to evaluate CPU usage savings.
%The second benchmark, called \emph{synchrobench}, is a well known test suite used to evaluate the effectiveness and efficieny
The tests were executed %on two different hardware platforms: a) an HP Proliant server equipped with 4 AMD Opteron 6168 for a total of 48 cores and 64GB of memory arranged in 8 NUMA nodes and b) 
on a ThinkMate GPX XT10-2260V4-4GPU equipped with 2 Intel Xeon E5-2640 v4 for a total of 20 cores equipped with 64GB of memory arranged in 2 NUMA nodes.

Figure \ref{fig:expeval} show all the results of the synthetic benchmark. In the left column we can find the throughputs in terms of critical section executed per second, while the center column shows CPU time wasted for synchronizing threads. Finally, the right column reports the ratio between the average throughtput of a given lock and the average of the optimal solution, namely the one that for each thread count has the maximum throughput achieved with the evaluated locks.  
The column denoted as PT-EXP is the mean between the values of PT-SPINLOCK and PT-MUTEX, representing the expected behaviour when a static choice (with uniform probability) between PT-SPINLOCK and PT-MUTEX has been made without knowing the actual workload and concurrency level.
If PT-EXP is higher than MUTLOCK, it is convinient taking the risk of an a-priori choice between the two pthread locks, otherwise MUTLOCK represent a better choice capable of ensuring an expected higher throughput. 

Figure \ref{fig:mltest1} shows the throughput while executing both critical sections and non-critical sections uniformly distributed between $0\mu s$ and $3.7\mu s$.
As expected, MCS has the highest throughput because it best fits the NUMA arrangement of memory when the thread count is lower than (or equal to) the number of cores. 
In fact, each thread spins on its own cache line and the thread holding the lock touches a single line (the one owned by the next in the FIFO order) for signaling the release. 
MUTLOCK has slightly lower throughput than PT-SPINLOCK, showing an overhead up to the 8\% for its management.
Conversely, PT-MUTEX (PT-ADAPTIVE-MUTEX) has a 25\% (12\%) drop of performance w.r.t. spin locks and shows its benefit only in case of time sharing, where going to sleep is a smart choice to reduce hardware contention.
As expected, MCS has a drop of performance in this scenario due to its FIFO semantics. 
The CPU usage devoted to synchronization is shown in Figure \ref{fig:mltest2}.
Here, we can see that our MUTLOCK consumes the same amount of CPU w.r.t. PT-SPINLOCK (and much less than MCS which is in trashing), conversely both mutexes reduce by one order of magnitude the CPU time.
This shows that mutexes are very efficient in terms of cpu usage, but this come with a price in terms of performance.
Figure \ref{fig:mltest10} confirms that spin locking is the best option with no-timesharing and PT-SPINLOCK has comparable performance w.r.t. PT-MUTEX in timesharing.
However, our MUTLOCK guarantees an higher average throughput than PT-EXP and almost optimal in case of timesharing.

In a second set of experiments we consider 
%evaluating locks behavior with
critical-section length uniformly distributed in $[0\mu s, 366\mu s)$ (Figure \ref{fig:mltest3}).
Here, we can see that spinning for a very long time is convenient only for low thread counts (up to 4).
Conversely, mutexes show their advantages having a maximum and stable throughput with thread counts higher than 4.
It follows that we are in a scenario where the hardware contention has a relevant impact on performance. 
While pure spin locking is fated to worse while increasing the thread count,
%Our solution shows its main limitation.
%In fact, even though 
MUTLOCK maintains a stable throughput, a bit lower than mutexes since it continues to keep 
%a higher number of threads spinning than mutex, has shown in Figure \ref{fig:mltest4},  
a few threads spinning to mask wake up latency. However, since critical sections are very long, such latencies are negligible and going to sleep allows to reduce hardware contention.
%In other words, our MUTLOCK is not able to keep into account two main factors while choosing the proper SWS: a) the CS %length and b) hardware contention.
Finally, MUTLOCK reduces the CPU time spent while synchronizing by an order of magnitude w.r.t. to spin locks for high thread counts (above 10) as shown in Figure \ref{fig:mltest4}.
Figure \ref{fig:mltest11} shows that this is a worst case scenario for our approach, since it has an average performance slightly lower than PT-EXP without timesharing.
This is reasonable since one of the main limitation of our approach is not considering the length of the critical section while sizing the spinning window. 
However, the proposed hybrid approach guarantees a loss bounded by the 8\% of the optimum, a threashold already crossed by pure spin locking approaches when running with 8 threads.

Short critical sections and non-critical section uniformly distributed in $[0\mu s, 366\mu s)$ 
lead to very low lock contention. 
%represents the case where locks are very low contented. 
In such a scenario all locks have similar performance (Figure \ref{fig:mltest5}) and CPU times (Figure \ref{fig:mltest6}), except for pure spin locks in case of time sharing. 
Consequently, there is almost no loss in adopting our MUTLOCK (Figure \ref{fig:mltest12}).

The last case, namely the one with both CS and NCS uniformly distributed in $[0\mu s, 366\mu s)$, resembles a scenario where critical sections are very long, but scarcely accessed.
%, thus reducing the effects of hardware contention.
On the one hand, this reduces the differences between pure spin locks and MUTLOCK as shown in Figure \ref{fig:mltest7} and, on the other hand, it exacerbates the benefits of making a controlled number of threads spinning in order to save CPU time and not paying wake up latency.
Clearly, the higher throughput than mutexes comes with a price in terms of CPU times (Figure \ref{fig:mltest8}).
In fact, our MUTLOCK consumes up to one order of magnitude of additional CPU time w.r.t mutexes in order to guarantee performance almost optimal.

To summarize, our approach allows to achieve the highest average performance with low contention and higher throughput than the expetation when a static choice between pthread spin lock and mutex has been made.
This makes our MUTLOCK a good candidate when operating in uncertain conditions because of an unpredictable (or difficult to be reasoned) workload and/or of a virtualized hardware.

We also tested our solution 
%Then, each lock has been adopted as synchronization mechanism in our open source 
within the open source share-everything Parallel Discrete Event Simulator (PDES) project\footnote{Available at \url{https://github.com/HPDCS}}. 
In this framework the simulation model is partitioned in Logical Processes (LPs) and its execution is guided by the occurance of discrete events handled by working threads mapped to specific CPU-cores. 
%Typically, such platforms adopt a speculative approach to enhance scalability, trying always to process events and %managing incosistencies due to causality violations via rollbacking the simulation to a safe state.
%Classical approaches makes LPs binded to computational units, namely worker threads (WTs), to favour data partitioning %and, consequently, they require mid/long term load balancing in order to all the computational power.
%Conversely, 
%The share-everything approach exploits multi-core machines by relying on a unique fully shared event pool, accessed by all the thread processing simulation events in critical sections.
%adopt only a very short-term binding where LPs are assigned to WTs only for the execution of a single event.
%This guarantees that computational power is always devised to process the highest priority events and reducing the %rollback probability.

As test-bed application we used the classical PHOLD benchmark \cite{Fuj90} configured with 1024 simulation objects and 16 or 20 worker threads.
%, where  Each LP schedules
%events for any other LP in the system, with an exponential timestamp increment. 
As usual for PHOLD, event processing leads to spending some CPU time, via a configurable busy loop emulating a given event granularity. In our experiments we initially set the loop to give rise to events with granularity varying in 25,50 and 100 microseconds, which can be considered from mid- to very-high-weight values.
In our PHOLD configuration we included 32 (greater than the number of cores) hot-spot simulation objects, towards which a given percentage of events are routed. This percentage has been set to 50\%. 
%It is known that PDES workloads with hot spots are difficult to manage since they might provide unbalance in case of %traditional PDES platforms relying in the binding between LPs and WTs. 
Figure \ref{fig:pdes} shows the speedups w.r.t the sequential execution and CPU times wasted for synchronization with the different lock implementations when running with 16 and 20 threads (with lower thread count results are very similar for different locks).
%for event granularity  $\mu$s.
The MUTLOCK allows the simulator to achieve the highest speed up showing that a static decision between spinning and sleeping is suboptimal for performance. 
This gain is observed in combinatijn with a significant reduction of the CPU time spent for synchronizaiton compared to 
PT-SPINLOCK.

%even in case not evaluated by lockbench. 

\vspace*{-0.2cm}

\section{Conclusions}
\vspace*{-0.1cm}

In this article, we have introduced the Mutable Locks, a locking mechanism based on the concept of spinning windowthat can control the number of threads enabled to spin in order to save CPU usage and to guarantee responsiveness while synchronizing threads. 
Finally, we have demostrated the validity of our  approach thanks  to an  extensive experimental evaluation in both synthetic and \linebreak \newpage \noindent real-world scenarios, showing that is capable of ensuring either higher performance or lower loss than evaluated adversaries.
As future work, we plan to study other approaches to resize the spinning window and to extend the states of waiting threads besides the classical spin/sleep ones, for example by introducing additional states where threads adopt a backoff time before attempting to acquire the lock or where CPU-cores are allowed to spin with a given frequency set by exploiting the Dynamic Voltage and Frequency Scaling capabilities of modern processors. 
This should allow to further increase the saving of computing power due to active waiting phases and to reduce hardware contention without sacrificing performance.

\end{document}